\documentstyle[12pt]{article}
\def\x {\stackrel {\textstyle \otimes}{,}}
\def\beq{\begin{equation}}
\def\eeq{\end{equation}}
\def\be{\begin{displaymath}}
\def\ee{\end{displaymath}}

\newcommand{\A}{\alpha}
\newcommand{\B}{\beta}
\newcommand{\C}{\gamma}
\newcommand{\sn}{{\rm sn\,}}
\newcommand{\cn}{{\rm cn\,}}
\newcommand{\dn}{{\rm dn\,}}

\catcode`\@=11

\font\teneus=eusm10 scaled \magstep1
\font\seveneus=eusm7 scaled \magstep1
\font\fiveeus=eusm5 scaled \magstep1

\newfam\eusfam
\textfont\eusfam=\teneus  \scriptfont\eusfam=\seveneus
  \scriptscriptfont\eusfam=\fiveeus

\def\hexnumber@#1{\ifnum#1<10 \number#1\else
 \ifnum#1=10 A\else\ifnum#1=11 B\else\ifnum#1=12 C\else
 \ifnum#1=13 D\else\ifnum#1=14 E\else\ifnum#1=15 F\fi\fi\fi\fi\fi\fi\fi}

\def\Cl{\ifmmode\let\next\Cl@\else
 \def\next{\errmessage{Use \string\Cl\space only in math mode}}\fi\next}
\def\Cl@#1{{\Cl@@{#1}}}
\def\Cl@@#1{\fam\eusfam#1}

\catcode`\@=\active

\catcode`\@=11

\font\teneuf=eufm10 scaled \magstep1
\font\seveneuf=eufm7 scaled \magstep1
\font\fiveeuf=eufm5 scaled \magstep1

\newfam\euffam
\textfont\euffam=\teneuf  \scriptfont\euffam=\seveneuf
  \scriptscriptfont\euffam=\fiveeuf

\def\hexnumber@#1{\ifnum#1<10 \number#1\else
 \ifnum#1=10 A\else\ifnum#1=11 B\else\ifnum#1=12 C\else
 \ifnum#1=13 D\else\ifnum#1=14 E\else\ifnum#1=15 F\fi\fi\fi\fi\fi\fi\fi}

\def\Got{\ifmmode\let\next\Got@\else
 \def\next{\errmessage{Use \string\Got\space only in math mode}}\fi\next}
\def\Got@#1{{\Got@@{#1}}}
\def\Got@@#1{\fam\euffam#1}

\catcode`\@=\active

\catcode`\@=11

\font\tenmsx=msxm10 scaled \magstep1
\font\sevenmsx=msxm7 scaled \magstep1
\font\fivemsx=msxm5 scaled \magstep1
\font\tenmsy=msym10 scaled \magstep1
\font\sevenmsy=msym7 scaled \magstep1
\font\fivemsy=msym5 scaled \magstep1
\newfam\msxfam
\newfam\msyfam
\textfont\msxfam=\tenmsx  \scriptfont\msxfam=\sevenmsx
  \scriptscriptfont\msxfam=\fivemsx
\textfont\msyfam=\tenmsy  \scriptfont\msyfam=\sevenmsy
  \scriptscriptfont\msyfam=\fivemsy
\def\hexnumber@#1{\ifnum#1<10 \number#1\else
 \ifnum#1=10 A\else\ifnum#1=11 B\else\ifnum#1=12 C\else
 \ifnum#1=13 D\else\ifnum#1=14 E\else\ifnum#1=15 F\fi\fi\fi\fi\fi\fi\fi}

\def\Bbb{\ifmmode\let\next\Bbb@\else
 \def\next{\errmessage{Use \string\Bbb\space only in math mode}}\fi\next}
\def\Bbb@#1{{\Bbb@@{#1}}}
\def\Bbb@@#1{\fam\msyfam#1}

\catcode`\@=\active

\title{
$R$-matrices for  the $n=2,3$  Elliptic Calogero-Moser Models \\
       }
\author{H. W. Braden  \quad \& \quad  T. Suzuki \\
\\
\normalsize
Department of Mathematics and Statistics,\\
\normalsize
University of Edinburgh,\\
\normalsize
Edinburgh, UK.
}

\begin{document}
\begin{titlepage}
\renewcommand{\thepage}{}
\maketitle
\vskip-9.5cm
\hskip11.4cm
Edinburgh-/92-93/03
\vskip.2cm
\hskip11.4cm
\sf hep-th/9312031  \rm
\vskip8.8cm

\begin{abstract}
The classical (dynamical) $R$-matrices for the 2- and  3-body Calogero-Moser
models with elliptic potentials are given.
The 3-body case has an interesting nontrivial structure that goes beyond
the known ansatz for momentum independent $R$-matrices.
The $R$-matrices  presented include the dynamical $R$-matrices of Avan
and Talon as  degenerate cases of the elliptic potential.
\end{abstract}
\vfill
\end{titlepage}
\renewcommand{\thepage}{\arabic{page}}

The elliptic Calogero-Moser model \cite{Cal75,Moser}
is a completely integrable  system
of $n$-particles on the line interacting via pairwise (so called type IV)
potentials $v(q_{ij})=a\wp(aq_{ij})$ with $q_{ij}=q_i-q_j$ \cite{OlP81}.
The function  $\wp$ is the Weierstrass elliptic function, an even
doubly-periodic  function with periods  $2\omega_1$ and $ 2\omega_2$;
the various degenerations of this function also give
 completely integrable  systems.
In the limit in which one
of the periods goes to infinity the potential reduces to  either
the  (type II) potential $v(q)=a\sp2/{\sinh(aq)}\sp2$ or the (type III)
potential $v(q)=a\sp2/{\sin(aq)}\sp2$.
When both periods go to infinity we obtain  the (type I)
potential $v(q)=q\sp{-2}$.
Although the Calogero-Moser model can be generalised to
arbitrary semisimple Lie algebras our attention in this letter will
be on the original Calogero-Moser model
which corresponds to the algebra $gl_n$.

The model has a Lax  pair formulation $i \dot{L} = [L, M]$.
The $L$ operator may be expressed as
\beq
L\equiv \sum_{\mu}L\sp\mu X_\mu = \sum_{i=1}\sp{n} p_i H_i +
i\sum_{\alpha\in \Phi}w_\alpha E_\alpha. \label{eq:laxl}
\eeq
Here $X_\mu=\{H_i, E_\A\}$ is a Cartan-Weyl basis for the algebra $gl_n$
with $\{H_i\}$ a basis for the Cartan subalgebra and $\{E_\A\}$ the
corresponding  set of step operators.
For the Lie algebra $gl_n$ the root system,  which
labels the step operators, may be written as
$\Phi=\{e_i-e_j,\; 1\le i\neq j\le n\}$,  with $e_i$ being an orthonomal
basis of ${\Bbb R}\sp{n}$. We may view the positions $(q_1,q_2,\ldots,q_n)$
as components of a vector  ${ q}$ in this vector space and so for
$\A=e_i-e_j$ we have $q_{ij}=\A\cdot q$.
It will be convenient to use the shorthand $f_\alpha$ for
the function that takes the value
$f(\alpha\cdot q)$  when evaluated at $q$.
The fuction $w_\A$ in (\ref{eq:laxl}) is
determined by the requirement that
the Lax equation reproduces the Hamiltonian equations of motion. This
means the function $w_\A$ should satisfy the following
addition formula,
\beq
w_\alpha w_\beta\sp\prime -w_\beta w_\alpha\sp\prime=
(z_\alpha-z_\beta) w_{\alpha+\beta}, \quad\quad{\rm where}\quad
z_\alpha(u) = {w_\alpha\sp{\prime\prime}(u)\over {2 w_\alpha(u)}}.
\label{eq:addition}
\eeq
The solutions  to this functional equation are given by
\beq
w_\alpha=
{\sigma(u-\alpha\cdot q)\over{\sigma(u)\sigma(\alpha\cdot q)}}
{e\sp{\zeta(u)\alpha\cdot q}}.
\label{eq:wfun}
\eeq
Here $\sigma(x)$ and $\zeta(x)=\sigma'(x)/\sigma(x)$ are Weierstrass's
sigma and zeta function.
The function $w_\A$  given in (\ref{eq:wfun})
depends not only on the coordinate
differences $q_{ij}=\A\cdot q$
but also on a spectral parameter $u$ \cite{Krich}.
If the spectral parameter $u$ is equal to one of $\omega_1$, $\omega_2$ or
$-\omega_1-\omega_2$ then $w_\A$ becomes $1/\sn \A,\, \cn \A/\sn \A$ or
$\dn\A/\sn\A$,  where $\sn\A, \cn\A$ and
$\dn\A$ are  the Jacobi elliptic functions.
In such cases $w_\A$ is an odd function, $i.e.$, $w_{-\A}=-w_\A$, and
the Lax operator $L$ is hermitian.

The Lax pair formulation immediately shows the functionally
independent  quantities $I_m\equiv {\rm Tr}\, L\sp m$  $(m\le n)$ to be
conserved.
Their Poisson commutativity, $\{I_m, I_k\}=0$, and hence the
complete integrability of the model, follows from
the existence of a ${\Got{g}}\otimes{\Got{g}}$-valued $R$-matix
\cite{STS,FT,BV90},
$R=R\sp{\mu\nu}X_\mu\otimes X_\nu$, satisfying,
\beq
\{L\x L\} = [R, L\otimes 1] - [R\sp{\pi}, 1\otimes L].
\label{eq:key}
\eeq
Here $R\sp\pi\equiv PRP\sp{-1}$ with $P$ being the permutation map:
$P(a\otimes b)P\sp{-1}=b\otimes a$.
Proving the  Poisson commutativity $\{I_m, I_k\}=0$, or equivalently
finding such an $R$-matrix, is one of the central tasks in proving the
integrability of a model.
Recently $R$-matrices for the Calogero-Moser models with types I, II, III,
and V potentials  subject to $w_{-\alpha}=-w_\alpha$  were
constructed by Avan and Talon \cite{AT},
while spectral parameter dependent $R$-matrices for the
type IV potential were independently given
by Sklyanin \cite{Sk}
and the authors \cite{BS}.
Moreover, we  also proved \cite[Theorem 4.6]{BS}
that when $n\ge4$  no momentum independent
$R$-matrices  exist for nondegenerate type IV potentials
satisfying $w_{-\A}=-w_\A$.
However, as mentioned in \cite{BS}, the Lie algebra consistency
equations  that forbid momentum independent
$R$-matrices for $n\ge4$ are trivially satisfied for $n=2, 3$.
In this letter we will present explicit $R$-matrices for the
$n=2, 3$ systems with potential satisfying $w_{-\alpha}=-w_\alpha$.
The structure found in the $n=3$ case is rather interesting
and provides an example of an $R$-matrix that goes beyond the known
ansatz.

Let us start the calculation of  the $R$-matrix.
To begin with, we decompose eq.(\ref{eq:key}) into  components with
respect to the basis $X_\mu\otimes X_\nu$.
Upon substituting $L$ into (\ref{eq:key}) and comparing
both sides, we obtain three equations,
\begin{eqnarray}
0 &=& \sum_\A (R\sp{\A j}\A_i - R\sp{\A i}\A_j) w_{-\A}, \\
-\A_i w'_\A&=& i\A\cdot p\, R\sp{\A i}+ \A\cdot R\sp i w_\A
      + \sum_\B (\B_i w_\B R\sp{-\B\A}+ w_{\A-\B}R\sp{\B i}c\sp\A_{\B\,\A-\B}),
\\
0 &=& \A\cdot R\sp\B w_\A - \B\cdot R\sp\A w_\B +i(\A\cdot p\,R\sp{\A\B}
-\B\cdot
        p\, R\sp{\B\A}) \\
{}&{}& +\sum_{\C\in\Phi}
 (R\sp{\C\B} c\sp\A_{\C\,\A-\C}w_{\A-\C}- R\sp{\C\A}c\sp\B_{\C\, \B-\C}
w_{\B-\C}).  \nonumber
\end{eqnarray}
The structure constant $c\sp\lambda_{\mu\,\nu}$ are defined by $[X_\mu, X_\nu]=
         c\sp\lambda_{\mu\,\nu}X_\lambda$ and
we have introduced the shorthand notation
$\A\cdot R\sp\B=R\sp{i\B}- R\sp{j\B}$ for $\A=e_i-e_j$.
The assumption that the $R$-matrix is momentum independent and that
$w_{-\A}=-w_\A$ greatly simplify these equations. We find
\begin{eqnarray}
&&R\sp{\A i}=0, \quad\; \quad
   R\sp{\A \B}=0 \quad{\rm if}\quad \A\neq\pm\B, \label{eq:one} \\
&&R\sp{\A \A}+R\sp{-\A -\A}=0, \quad R\sp{\A -\A}+R\sp{-\A \A}=0, \quad
 R\sp{\A \A} + R\sp{-\A \A}
    =-\frac{w'_\A}{w_\A},  \label{eq:two}
\end{eqnarray}
and
\beq
 \A\cdot R\sp\B w_\A - \B\cdot R\sp\A w_\B
          = c\sp\B_{\A\C}(R\sp{\A\A} +  R\sp{\B\B}) w_{\C} +
            c\sp\B_{-\A{\gamma\sp\prime}}(R\sp{-\A\A}  +  R\sp{-\B\B})
            w_{\gamma\sp\prime}.
\label{eq:three}
\eeq
Also $R\sp{ij}=P\sp{ij}$, where $P\sp{ij}$ is a matrix such that
$\alpha\cdot P\sp{j}=0\ \forall \alpha\in\Phi$ \cite[Lemma 4.1]{BS}.
Note that at most one term on the  righthand side of
(\ref{eq:three}) is nonzero. If,
for example, $\C=\B-\A\in\Phi$ we have only the first term nonvanishing
and so
$\A\cdot R\sp\B w_\A - \B\cdot R\sp\A w_\B=c\sp\B_{\A\C}(R\sp{\A\A}+R\sp{\B\B})
w_{\C}$.
However (as $-\B,-\C\in\Phi$) we also know that $-\B=-\A-\C\in\Phi$
and so we obtain from the second term (after utilizing
$c\sp{-\C}_{-\A-\B}=- c\sp{\C}_{\A\B}$) that
$\A\cdot R\sp{-\B}w_\A - \B\cdot R\sp\A
w_\B=c\sp\B_{\A\C}(R\sp{-\A\A}+R\sp{\B-\B}) w_{\C}$.
These exemplify some of the consistency requirements mentioned earlier
that forbid (nodegenerate) solutions  for $n\ge4$.
We shall now solve them explicitly for $n=2,3$.
\newline
\underline{The case $n=2$}$:\quad$
In this case the roots are $\pm\A \ (\A=e_1-e_2)$  and the
structure constants on the right hand side of (\ref{eq:three})
vanish giving  the one equation
\beq
\A\cdot(R\sp{\A}-R\sp{-\A})=0.
\label{eq:n2a}
\eeq
The vanishing of the structure constants means equations (\ref{eq:two})
and (\ref{eq:n2a}) decouple and we may solve  (\ref{eq:two})
in terms of an arbitrary function $f$ by
\beq
R\sp{\A\A}=-R\sp{-\A-\A}=f \quad\quad{\rm and}\quad\quad
R\sp{\A-\A}=-R\sp{-\A\A}=f+\frac{w'_\A}{w_\A}.
\eeq
Similarly we may solve (\ref{eq:n2a}) in terms of  arbitrary
functions $g_1,g_2$ and $h$ by
\beq
R\sp{i\A}=g_i \quad\quad{\rm and}\quad\quad R\sp{i-\A}=g_i+h.
\eeq
Finally $P\sp{ij}=({\epsilon_1\atop\epsilon_1}{\epsilon_2\atop\epsilon_2})$
satisfies
$\alpha\cdot P\sp{j}=0$. In matrix form this means the $R$-matrix looks
like
\beq
\pmatrix{\epsilon_1 &g_1   &0                      &f   \cr
         g_1+h   &\epsilon_2   &f+\frac{w'_\A}{w_\A}&0      \cr
         0       &-f-\frac{w'_\A}{w_\A}&\epsilon_1&g_2  \cr
         -f   &0     &g_2+h            &\epsilon_2          \cr }.
\eeq
By choosing $\epsilon_1=\epsilon_2=f=0$ and
$g_1=g_2=-h/2=-w_\A$ we obtain an
$R$-matrix in the form of Avan and Talon's \cite[eq.(17)]{AT}.
\newline
\underline{The case $n=3$}:$\quad$
Corresponding to any root $\A\in\Phi$ there are now precisely two
roots $\C_{1,2}$ such that $\A+\C_{1,2}\in\Phi$. Indeed $\C_1+\C_2=-\A$.
This peculiarity of $n=3$ allows us to make the following ansatz for
$R\sp{\A\A}$:
\beq
R\sp{\A\A}=-{1\over2}\left( \frac{w'_\A}{w_\A} + \frac{w'_{\C_1}}{w_{\C_1}}
+\frac{w_\A w_{\C_1}}{w_{\A+\C_1}} \right)
          =-{1\over2}\left( \frac{w'_\A}{w_\A} + \frac{w'_{\C_2}}{w_{\C_2}}
+\frac{w_\A w_{\C_2}}{w_{\A+\C_2}} \right).
\label{eq:Raa}
\eeq
The  righthand equality between the two expressions involving the
different roots $\C_{1,2}$ follows from the addition formula
(\ref{eq:addition}) together with the identity
$z_\A=\frac{3}{2}\wp(u)+w_\A\sp2$ for
$u\in\{\omega_1,\omega_2,-\omega_1-\omega_2\}$.
Thus $R\sp{\A\A}$ is well defined\footnote{
In the notation of \cite{BS} this choice corresponds to
${\cal A}=-{3\over2}\wp(u)$.}
and depends only on the choice of $\A$.
Further $R\sp{\A\A}+R\sp{-\A-\A}=0$ and  the remaining equations of
(\ref{eq:two}) are solved upon  setting
\beq
R\sp{-\A\A}={1\over2}\left(-\frac{w'_\A}{w_\A} + \frac{w'_{\C_1}}{w_{\C_1}}
+\frac{w_\A w_{\C_1}}{w_{\A+\C_1}} \right)
          ={1\over2}\left(-\frac{w'_\A}{w_\A} + \frac{w'_{\C_2}}{w_{\C_2}}
+\frac{w_\A w_{\C_2}}{w_{\A+\C_2}} \right).
\label{eq:Rmaa}
\eeq
Our ansatz has the following two important features: if
$\B=\A+\C\in\Phi$ (for some $\C$), then $R\sp{\A\A}+R\sp{\B\B}=0$
while $R\sp{-\A\A}+R\sp{\B-\B}=w_\A w_\B /w_\C$.
Note this means the first term on the righthand side of (\ref{eq:three}) will
{\it always} vanish. This feature appears to hold for all of the $R$-matrix
ansatz for Calogero-Moser related models we know of; it is
usually achieved by setting $R\sp{\A\A}=0$.
Utilising these features means (\ref{eq:three}) simplifies to
\beq
 \A\cdot\frac{ R\sp{-\B}} {w_{-\B}} + \B\cdot \frac{R\sp\A}{w_\A}
          = -c\sp\B_{\A\C} .
\label{eq:four}
\eeq
This equation also arises (perhaps implicitly) in all of the $R$-matrix
ansatz we know of. For $gl_n$ a solution to the equation
\beq
\A\cdot A\sp{-\B}+\B \cdot A\sp\A=c\sp\B_{\A\C}
\label{eq:five}
\eeq
is given by $A\sp{i\A}=|e_i\cdot\A|/2$ and so $R\sp{i\A}=-|e_i\cdot\A|w_\A/2$
yields a possible solution to (\ref{eq:four}). Actually  (\ref{eq:five})
has  a more general solution. Because the vector $(1,1,1)$ is orthogonal
to every root then we may add a multiple of this to every solution
of (\ref{eq:five}). Thus $R\sp{i\A}=-|e_i\cdot\A|w_\A/2 +f\sp\A$ is also a
solution for arbitrary functions $f\sp\A$ (one for each root).
Further $A\sp{i\A}= \A_i f\sp{i}$
will satisfy the homogeneous
part of (\ref{eq:five}) for arbitrary functions $f\sp{i}$ and so we
obtain the solution
\beq
R\sp{i\A}=-|e_i\cdot\A|w_\A/2 +f\sp\A+\A_i f\sp{i}w_\A
\label{eq:Ria}
\eeq
in terms of the nine arbitrary functions $f\sp\A,f\sp{i}$. A similar
solution holds for arbitrary $n$.
Observe that (\ref{eq:four}) contains $9$ constraints on the
$18$ components $R\sp{i\A}$, six of them corresponding to three
different roots which sum to zero and three coming from $\alpha=\beta$
for $\alpha$ a positive root. This accounts for the nine arbitrary
functions in our solution.

Finally
$P\sp{ij}=\big({{\epsilon_1}\atop{\epsilon_1\atop\epsilon_1}}
           {\epsilon_2\atop{\epsilon_2\atop\epsilon_2}}
           {\epsilon_3\atop{\epsilon_3\atop\epsilon_3}}\big)$
satisfies
$\alpha\cdot P\sp{j}=0$. This means $R$-matrix
takes the form  given in Table 1, where for $\alpha=e_i-e_j$ we have set
$w_\A\equiv w_{ij}$, $f\sp\A\equiv f\sp{ij}$ and $R\sp{\A\A}={\cal R}_{ij}$.
We note that  as the elliptic potential degenerates,
$\lim_{k\to0}{R\sp{\A\A}}=0$. Therefore the $R$-matrix obtained by
setting $\epsilon_i=f\sp{i}=f\sp\A=0$ yields the solution of
Avan and Talon for the type I, II and II potentials.

\vskip 0.5in

To conclude.
In this letter, we have presented general momentum independent
$R$-matrices for  the
$n=2$ and $n=3$ Calgero-Moser systems with elliptic (type IV) potentials
of Jacobi type, {\em i.e.}, $w(q)=-w(-q)$.
These $R$-matrices reduce to the known solutions given by
Avan and Talon for the potentials of the type I, II and III
and are the first $R$-matrices known for these two particular
models.
The $R$-matrices found here have $R\sp{\A\A}\neq0$ for
the  general elliptic potential.
This feature is new: the $R$-matrices normally associated
with the Calogero-Moser models have $R\sp{\A\A}=0$.
The nontrivial ansatz (\ref{eq:Raa}) makes full use of the
$n=3$ root space geometry and it is the particularly simple
structure of the $n=2,3$ root spaces that allows a  momentum independent
solution.

\vskip 0.5in
T.\,S. acknowledges financial support from both the
Daiwa Anglo-Japanese Foundation and Fuju-kai Foundation.

\end{document}